\documentclass{article}
\usepackage[utf8]{inputenc}

\usepackage{hyperref}
\usepackage{color}
\usepackage{authblk}
\usepackage{float}
\usepackage{graphicx}
\usepackage{amssymb,amsmath}
\usepackage[english]{babel}
\usepackage[nottoc]{tocbibind}
\usepackage{geometry}
\usepackage[numbers,sort&compress]{natbib}

\title{Advancing Superconducting Magnet Diagnostics\\for Future Colliders}

\author[1,*]{M. Marchevsky}
\author[1]{R. Teyber}
\author[1]{G. S. Lee}
\author[1]{M. Turqueti}
\author[2]{\\M. Baldini}
\author[2,3]{E. Barzi}
\author[2]{J. DiMarco}
\author[2]{S. Krave}
\author[2]{V. Marinozzi}
\author[2]{S. Stoynev}
\author[4]{\\P. Joshi}
\author[4]{J. Muratore}
\author[5]{D. Davis}
\affil[1]{\it{Lawrence Berkeley National Laboratory}}
\affil[2]{\it{Fermi National Accelerator Laboratory}}
\affil[3]{\it{Ohio State University}}
\affil[4]{\it{Brookhaven National Laboratory}}
\affil[5]{\it{National High Magnetic Field Laboratory}}
\affil[*]{\url{mmartchevskii@lbl.gov}}

\date{March 15, 2022}

\begin{document}

\maketitle
\section*{Abstract}

Future colliders will operate at increasingly high magnetic fields pushing limits of electromagnetic and mechanical stress on the conductor \cite{1-MM}. Understanding factors affecting superconducting (SC) magnet performance in challenging conditions of high mechanical stress and cryogenic temperatures is only possible with the use of advanced magnet diagnostics. Diagnostics provide a unique observation window into mechanical and electromagnetic processes associated with magnet operation, and give essential feedback to magnet design, simulations and material research activities. Development of novel diagnostic capabilities is therefore an integral part of next-generation magnet development. In this paper, we summarize diagnostics development needs from a prospective of the US Magnet Development Program (MDP), and define main research directions that could shape this field in the near future.

\section*{Introduction}

Development of high-field accelerator magnets is a complex task aiming at a gradual improvement of magnet performance through understanding of causes of performance limitations. Diagnostic instrumentation and data analysis are thus playing a crucial role in this development process, improving efficiency and enabling better outcomes in terms of magnet performance and reliability of operation. Several key questions can be identified as targets for diagnostics development, aiming at the existing challenges in Nb$_{3}$Sn magnets, high-temperature superconductor magnets and diagnostics instrumentation itself.

For Nb$_{3}$Sn-based magnets \cite{2-MM,3-MM} a well-known long-standing problem is training: a process of gradually improving a maximal current the superconducting magnet can sustain with successive energizing up to a quench.  Upon repetitive quenching, the limiting current would gradually increase and eventually saturate at a plateau defining the ultimate magnet performance. Since dozens of repetitive quenching test are typically needed to reach the design current with some safety margin, training is a costly and time-consuming procedure that every newly constructed magnet has to undergo prior to its intended use. Closely connected to training is the problem of mechanical memory (or lack thereof) which represents an ability of a magnet to sustain its performance level achieved with training with subsequent thermal cycling. Finally, magnet operation at 15+ T field level translates into Lorentz force stresses of ~200 MPa at the brittle conductor which can then experience irreversible damage. Detecting and mitigating such damage is yet another important task for diagnostics. Training and mechanical memory are believed to be closely connected to various mechanical flaws such as cracks in epoxy impregnation or interfacial de-lamination. Energy releases associated with stress-driven formation of these flaws under mechanical stress are causing premature quenching while gradual “clearing” of these transients upon ramping up and quenching gives rise to the training phenomenon. To summarize, the following questions are standing:

\begin{itemize}
\item How do we resolve and properly classify mechanical and electromagnetic disturbances in SC Nb$_{3}$Sn magnets and understand the physics of the training process? 
\item How do we non-invasively localize weak points and interfaces where mechanical disturbances causing premature quenching are taking place? Can we perhaps manipulate those interfaces in situ to improve magnet performance? 
\item 	How do we detect and localize stress-driven conductor degradation?
\end{itemize}

High-temperature superconductors (HTS) are expected to have a major impact on accelerator magnets enabling dipole field strength above 15 T, in either hybrid (Nb$_{3}$Sn – HTS) or HTS-only configurations. One important problem that needs to be solved for HTS-based magnets is quench detection. Due to a large enthalpy margin of HTS conductors, normal zones propagate 1-2 orders of magnitude slower in HTS conductors than in low-temperature superconductor (LTS) conductors such as Nb$_{3}$Sn. As a result, voltage-based quench detection is of limited use for HTS since local hot spot temperature may rise above safe levels well before any voltage is detected. Various non-voltage techniques are presently under development to improve sensitivity and redundancy of quench detection for HTS magnets; some of those methods could allow localization of hot spots in addition to detection, providing valuable feedback to magnet engineers \cite{24-MM}. As non-propagating hot spots can form in HTS conductor well before quenching, a new paradigm for HTS magnet operation has recently emerged aiming at early detection of such hot spots and avoiding quenching altogether by safely ramping down the magnet current. Quench protection is yet another important part of HTS magnet operation. Efficient protection requires a good understanding of HTS coils in terms of current sharing and inductive coupling between individual conductors. It is because localized defects in HTS conductors can be potentially mitigated through efficient current sharing, allowing current to “bypass” defective regions and prevent thermal damage to the conductor. In addition to preexisting conductor defects, various mechanical flaws such as crack formation in the superconducting layer or filament as well as de-lamination of HTS layer from substrate in ReBCO tapes can occur over time in magnets operating under high mechanical stress. Timely detection and localization of those conductor flaws can ensure reliable magnet operation. In summary for HTS magnets critical questions are:

\begin{itemize}

\item How do we achieve a reliable and minimally invasive quench detection and localization capability for HTS \cite{4-MM,5-MM,6-MM,7-MM} and hybrid HTS/LTS \cite{8-MM} magnets? 
\item Can we practically realize a new paradigm of HTS magnet operation where quenching can be avoided altogether through early detection?
\item	How do we resolve current sharing patterns and stress-driven defect accumulation in HTS coils and cables to ensure their long-term operational stability and quenching resilience?

\end{itemize}

Diagnostic instrumentation is constantly evolving to provide better answers to the above questions for both LTS and HTS. A broad spectrum of novel diagnostic approaches is already being explored to address some of these critical questions by the multi-lab Diagnostics Working Group collaborating in the framework of the U.S. Magnet Development Program \cite{9-MM}. Acoustic instrumentation leveraging advances in piezo-sensors and compact cryogenic amplifiers allow for non-invasive measurements of mechanical energy release in magnets, 3D triangulation of mechanical disturbances and quenches in complex magnet systems with an accuracy of few centimeters \cite{10-MM}. Magnetic quench antennas \cite{12-MM} are used for localization of quenches and electromagnetic disturbances. Hall sensor arrays \cite{13-MM} enable mapping of current redistribution and quench detection in multi-conductor cables and coils. Diffuse-field ultrasonic methods allows for a non-invasive detection of hot spots in HTS coils and conductors \cite{11-MM} and structural integrity monitoring. Distributed sensing allowing for continuous monitoring of various physical quantities along the magnet structural elements and conductor path is the next frontier in magnet instrumentation, and can be implemented with fiber-optic \cite{14-MM}, radio-frequency (RF) or ultrasonic technologies. Capacitive sensing \cite{15-MM} are being actively explored for sensitive detection of temperature rise and quench detection in HTS coils immersed in liquid cryogen. Cryogenic electronic circuits and specialized data processing hardware based on field-programmable gate arrays (FPGA) enable new diagnostic measurement capabilities. As the amount of diagnostic data keeps increasing, data analysis techniques increasingly rely on new approaches such as machine learning, using approaches like unsupervised learning for identification of mechanical transient events, electrical disturbances or toward real-time quench detection.
Next, we review our plans for  developments of novel sensor hardware, electronics and data analysis techniques towards real-time, non-invasive monitoring of LTS, HTS and hybrid magnets. The end goal of such development would be the application of this advanced diagnostic system to existing and future accelerator magnets and magnet test facilities.

\section*{Diagnostics development}

Here are practical approaches we are going to take in our diagnostics development:

\begin{itemize}
\item Non-destructively and non-invasively localize the position of the quench including insulation failure and surface failure/delamination of the magnet.
\item	Analyze the electric circuit characteristics of HTS magnets and it's potential use in diagnosis technology.
\item	Establish fiber-optic based diagnostic capabilities through the use of Fiber Bragg Grating (FBG) and Rayleigh scattering-based sensors to measure elastic deformations, localize hot spots (especially in HTS magnets) and probe mechanical disturbances in SC cables \cite{16-MM}.
\item	Improve accuracy of voltage, magnetic and acoustic-based diagnostics through calibration using distributed spot heater and piezo-transducer arrays.
\item	Bring magnetic diagnostics to the next level through development and use of flexible multi-element quench antennas, large-scale Hall sensor arrays and non-rotating field quality probes, aiming at understanding electromagnetic instabilities in LTS magnets and imaging current-sharing patterns in superconducting cables and HTS magnet coils. Develop new algorithms for current flow reconstruction and disturbance localization.
\item	Design and conduct innovative small-scale experiments to probe training behavior and energy release in impregnated cables under similar loads as magnets \cite{17-MM,18-MM}. 
\item	Explore new methods for reliable and robust quench detection and localization for HTS magnets and hybrid LTS/HTS magnets.
\item	Use diffuse field ultrasonic techniques to enable targeted delivery of vibrational excitation to the conductor, for a non-invasive structural local probing of SC coils and mitigation of their training behavior.
\item	Apply machine learning and deep learning approaches to process diagnostic data and identify real-time predictors of magnet quenching.
\item	Develop cryogenic digital and analog electronics to facilitate, simplify and improve reliability of diagnostic instrumentation by enabling pre-processing of magnet diagnostic data in the cryogenic environment.

\end{itemize}

Below we discuss these approaches in more detail.

\clearpage
\subsection*{Acoustic emission diagnostics}

Premature quenching in superconducting magnets is most often associated with transient releases of mechanical energy such as crack formation, interface de-bonding and conductor motion. It is also believed that mechanical memory in magnets associated with training can be “stored” in the structural defects forming with those transients. Because individual instances of the transient events take place deep inside magnet coil windings and take only a fraction of a millisecond in most cases, they are difficult to access with classic diagnostic techniques such as strain, temperature or voltage monitoring. However, being of “transient” nature, these events generate elastic waves (acoustic emissions) in the magnet structure. That makes acoustic emission (AE) monitoring a very useful tool for quantifying transient energy releases, localizing their origins and learning about physical mechanisms of training and mechanical memory. Acoustic emission diagnostics is one of a few techniques available that enables simple, inexpensive and non-invasive access to transient micro-mechanics of superconducting magnets\cite{19-MM}.
Acoustic methods were shown to:
\begin{itemize}
\item	Spatially localize mechanical transients and quench locations using time of flight techniques \cite{10-MM,20-MM, 21-MM}
\item	Measure local energy release in those transients by relying on the known calibration methods \cite {10-MM}
\item	Record transient acoustic waveforms with a high temporal resolution and apply advanced spectral analysis techniques based on deep learning to classify various mechanical event types \cite{22-MM}
\end{itemize}

\begin{figure}[h]
		\centering
		\includegraphics[width=5in]{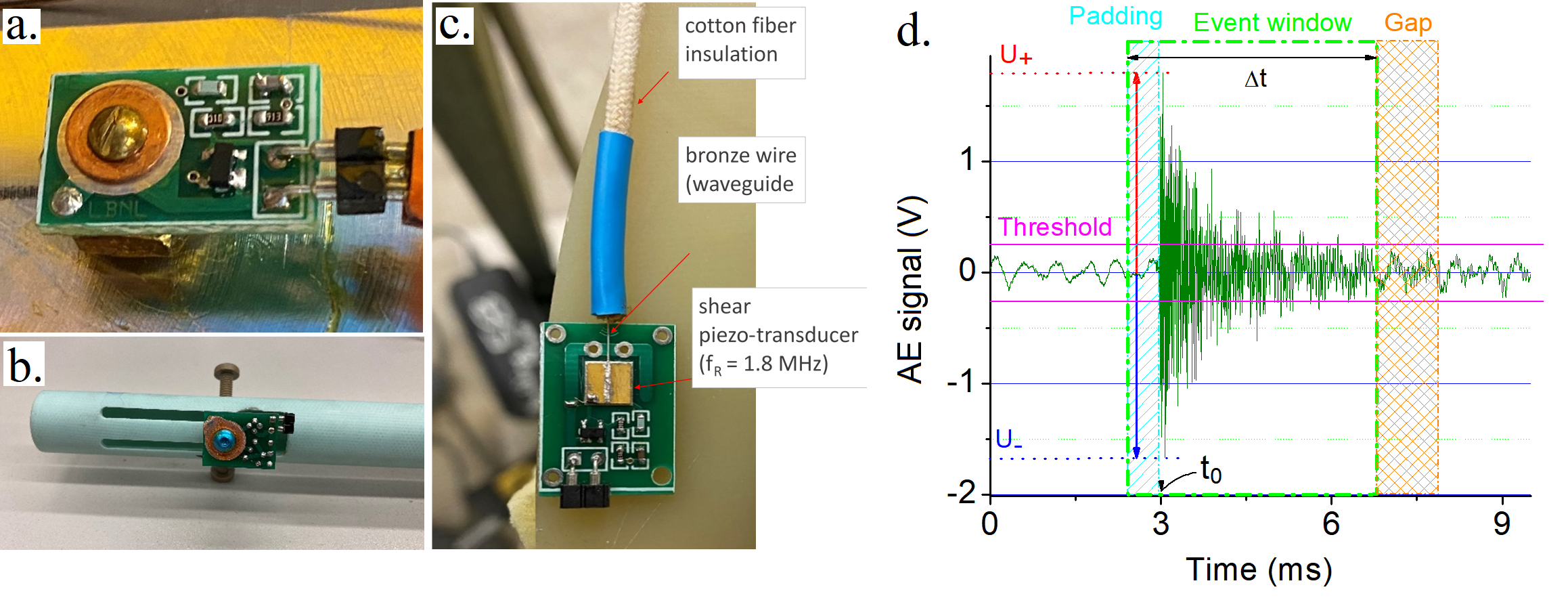}	
		\caption{(a) Cryogenic acoustic emission sensor developed at LBNL for direct installation on the magnet outer surface using a mounting screw (b) A G-10 cantilever mount for installing acoustic sensors in the magnet bore (c) Next generation ``waveguide" acoustic emission sensor.(d) Typical AE signal measured on a superconducting magnet.}
		\label{Fig:MM_AEsensors}
	\end{figure}
	
Cryogenic acoustic emission sensors (Fig.~\ref{Fig:MM_AEsensors}) having ~0-300 kHz bandwidth as well as various supporting structures for installation in the magnet bore were developed at Lawrence Berkeley National Laboratory (LBNL) for the US Magnet Development Program and have been instrumental for magnet testing at DOE labs. Our future AE hardware development will focus on improving sensitivity and fidelity of acoustic emission signals by picking up acoustic emissions from specific locations within the magnet using acoustic waveguides made out of metal wires with soft insulation and shear-mode piezoelectric transducers integrated within the printed circuit board of the cryogenic amplifier. These novel sensor designs enable bandwidths in excess of 1 MHz as well as high fidelity of the acoustic signals. The latter is important for improving localization accuracy of AE disturbances and classification transient events using machine learning methods.
In addition to passive AE monitoring, active acoustic diagnostics are in development where propagation of ultrasound emitted by an active transducer in the magnet structure is monitored. This approach allows for real-time detection of interface contacts between structural magnet elements in the process of current ramping and training by monitoring diffuse ultrasound waveforms \cite{23-MM}. A promising variation of this technique is acoustic reflectometry where wave propagation can be confined and guided within a single structural element with minimal scattering. Acoustic reflectometry may be potentially used to localize mechanical flaws and locations of strain and temperature variation within the magnet.

\subsection*{Quench Antennas}

Quench antennas continue to provide high-sensitivity, low-noise, low-cost, and non-invasive measurements of the quench disturbance spectrum in Nb$_{3}$Sn magnets \cite{RT-leroy, RT-ogitsu, RT-willeringQA, RT-dimarcoQA}. Design and fabrication have standardized around PCB technology (primarily flex, but also rigid), with the possibility of proximate mounting near the conductor within the magnet during assembly.  Hardware efforts should focus on improving the spatial resolution of activity and quench localization. This can be accomplished through grids of intersecting windings \cite{RT-stoyanQA} on multi-layer PCBs, and even creating small pixelated sensors, especially for use in areas of complex magnet cable geometry (e.g. in the magnet ends). The extensive scaling of measurement channels requires innovative electronics solutions around multiplexing and real-time mesh scaling to local measurements. More sophisticated geometries may offer the possibility of encoding unique response signatures for quench localization over a larger area for a single channel; these may require mapping of the array response coupled with machine-learning-assisted decoding. Hardware efforts should continue to relate measured disturbances back to field harmonics to help interpret redistribution phenomena in the magnet sense. Quench antenna analysis should be aimed not only at detecting quenches, but also at identifying the underlying disturbance mechanisms \cite{RT-keijzer} and monitoring the field harmonics in real time; this will require combined efforts of high-fidelity quench modeling as well as sub-scale and model tests. The combined hardware and analysis efforts outlined above should be explored in the context of upcoming hybrid LTS-HTS insert magnet tests, where there is a demand to sense mechanical and electromagnetic coupling between magnets.

 	\begin{figure}[H]
		\centering
		\includegraphics[width=5in]{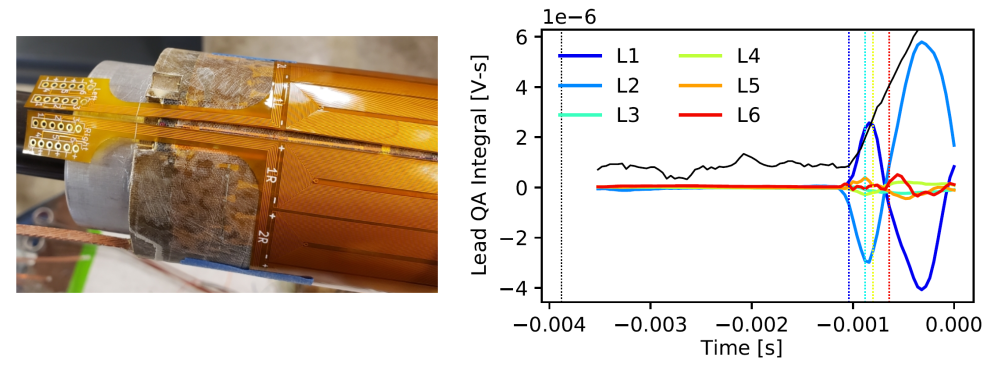}	
		\caption{(Left) Flexible quench antenna wrapped around Canted Cosine Theta (CCT) inner layer, and (Right) example integrated bucked voltages showing quench initiation around 30 degrees below pole (L1-L2).}
		\label{Fig:teyber_QA}
	\end{figure}

\subsection*{Distributed Spot Heater and Piezo-Transducer arrays}

The intricacies of quench development and current redistribution in Nb$_{3}$Sn magnets are still not mastered \cite{1-SS}, creating a need for sub-scale experiments and representative cable tests with well-controlled conductor conditions and a host of versatile instrumentation. Such experiments will improve understanding of natural quenches occurring in state-of-the art accelerator magnets. Of particular interest is a development program with distributed energy sources (e.g. spot heaters, piezo transducers) representing an array capable of emulating various conditions pertinent to quenches. The single elements could act on a level as low as a single strand and, acting in concert, be able to synthesize arbitrary conditions for exploration. This could enable new opportunities for bench-marking quench simulations, and both induced and natural quenches in the same coil and conductor area could be directly compared. Such an investment in targeted quench characterization would play an important role in both diagnostics development and calibration, as well as in improving understanding towards the driving LTS questions outlined above.

\subsection*{Current Distribution Monitoring and Hall Sensor Arrays}

\begin{figure}[H]
		\centering
		\includegraphics[width=0.9\textwidth]{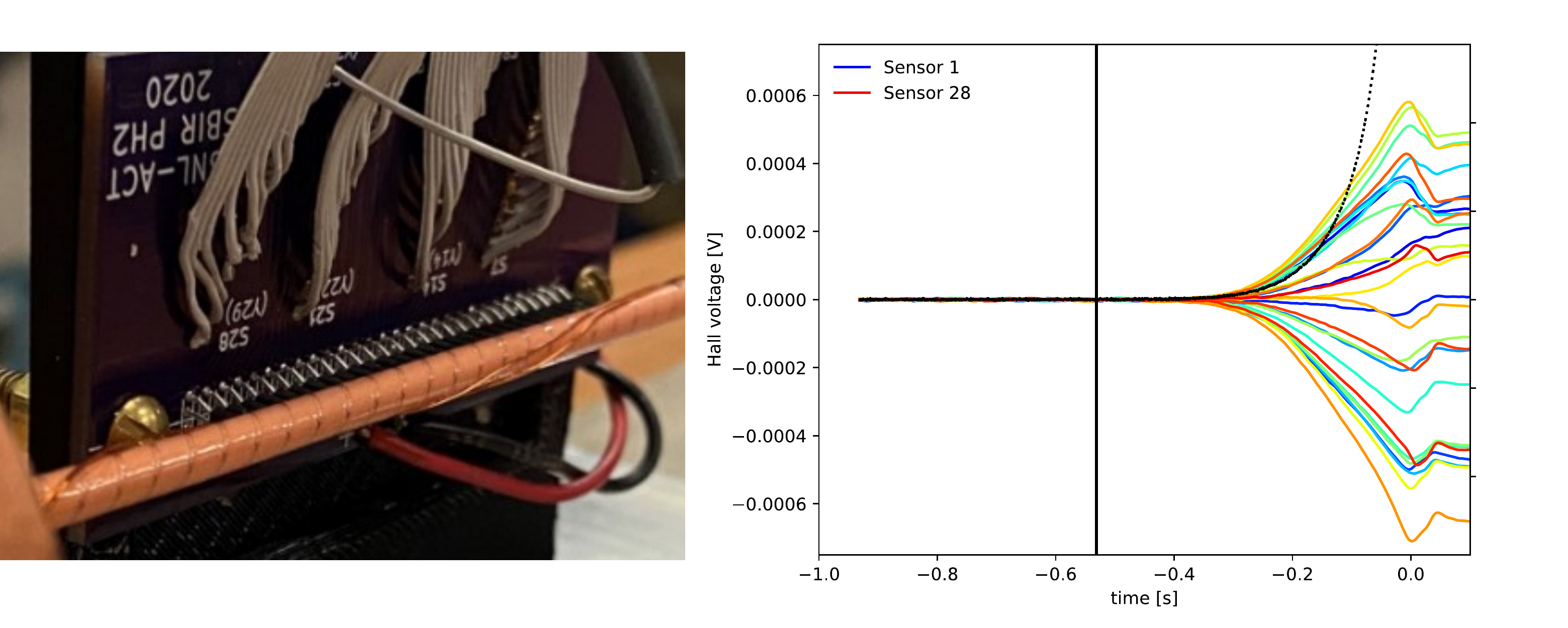}		
		\caption{(Left) Solenoidal Hall probe array near CORC termination, and (Right) Hall probe responses (colored) vs. sample voltage (black dotted line) after firing a quench heater (initiated at black vertical line). Current sensing can be an effective technique to monitor HTS magnets.}
		\label{Fig:teyber_1}
	\end{figure}

    The need to identify and predict current sharing patterns drives a number of diagnostics opportunities for HTS magnet development \cite{RT-willering, RT-takayasu}. Methods to determine the spatial distribution of HTS cable performance before winding should be developed, including techniques to extract performance parameters related to termination resistances, contact resistances, spatial distribution of critical currents and inductance matrices. In addition to identifying defects, such knowledge facilitates real-time monitoring and predictive operation where improved high-speed modeling tools capable of near-real-time prediction are desired; fault triggers can be generated when magnet measurements depart from model predictions. This "hardware-in-loop" magnet operation scheme relies on continued improvements in diagnostics aimed at probing current distributions in real time \cite{RT-marchevsky, RT-weiss, 13-MM, RT_ilyin}, which can be considered for both inter-tape redistribution in single cables (Fig.~\ref{Fig:teyber_1}) and for inter-cable redistribution in cable bundles (e.g. Ref.~\cite{RT-weiss}). The simplest cases pertain to cable configurations with poor current sharing (i.e. bundles of HTS cables for accelerator magnets); temporal knowledge of inter-cable current distributions near terminals have already been used in least-squares parameter extractions to monitor magnet operation. In this case, continued progress towards current sensing and efficient current recreation from stabilized Biot-Savart matrix inversion are desired. This is more complicated to implement non-invasively in cables with efficient current sharing. While Hall probe arrays can be used as a proxy for inter-tape current redistribution, high-speed \cite{RT-martinez} and high-fidelity \cite{RT-nugteren} HTS cable models along with improved cable characterization methods will increasingly play a role.

\subsection*{Non-invasive surface/insulation failure detection and localization using electrical reflectometry (RF reflectometry)}

\begin{figure}[H]
		\centering
		
		\includegraphics[width=0.45\textwidth, height=4cm]{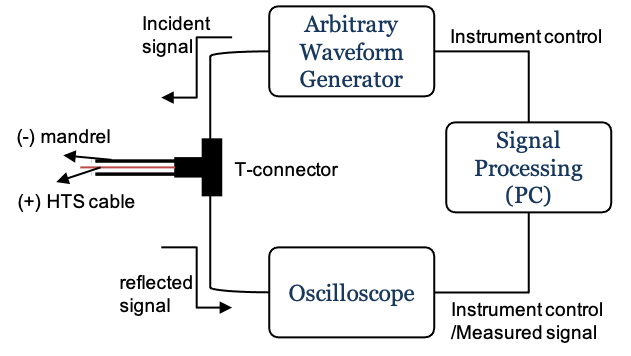}		
		\includegraphics[width=0.45\textwidth, height=4cm]{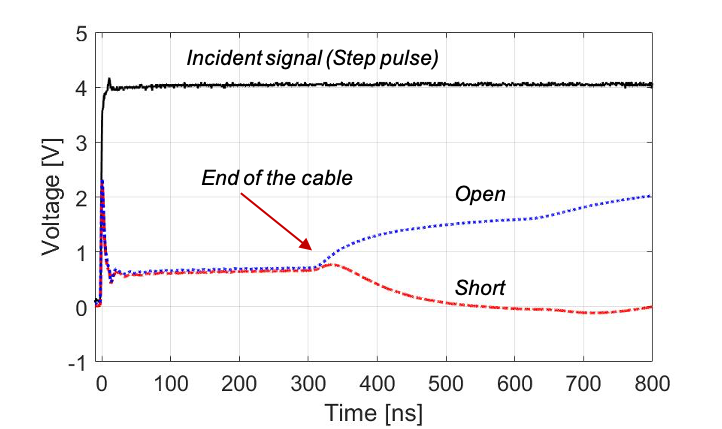}
	
		\caption{(Left) Electrical reflectometry measurement setup and (Right) typical time-domain reflectometry response from CCT subscale.}
		\label{Fig:GS_1}
	\end{figure}

For the electrical insulation of LTS and HTS magnets, epoxy resin, polyolefin-based resin, and paraffin wax are typically used. Furthermore, new development of epoxies and impregnation have been studied to improve the performance of magnets \cite{1-GS,2-GS}. However, unexpected defects on the surface or insulation can still be present as a result of problematic manufacturing processes, incomplete installation procedures, and quench. In order to protect against operational failures of magnets, methods to determine the location of surface/insulation failure should be developed including the method to classify the failure type. Electrical reflectometry can be used to non-destructively and non-invasively localize and identify the failures of magnets. Electrical reflectometry applies an electrical RF signal to the cable and measures the reflected signal at the point of impedance discontinuity as shown in Fig.\ref{Fig:GS_1}, regardless of cable length and insulation type. The impedance discontinuity can include insulation failure, structure failure, and superconducting cable resistance change. The shape of incident signals includes on pulse, rectangular, sinusoidal, and Gaussian enveloped chirp signals. Depending on the analysis method, the results of reflectometry is analyzed in the time domain, frequency domain, and time-frequency domain \cite{3-GS,4-GS,5-GS}.
Hardware efforts should focus on the ``distortion-free” transmission of high-frequency signal (more than 50 MHz). One of the challenging issues to accomplish the transmission is an impedance mismatching at the interface between the probe and the superconducting cable. The impedance matching circuit that minimizes signal reflection and loss through analysis of the magnet's matching network is required. Another issue is forming the transmission line of electrical signals with a single superconducting cable. This can be solved through additional accessories for micro-strip, co-axial, and two wire transmission line configurations. These additional accessories should be considered in the magnet design.

\subsection*{Capacitance Monitoring}

Most superconducting magnet systems operate in liquid nitrogen or helium, which have a drastic change in the dielectric properties as they evaporate; the relative permittivity drops approximately 40\% for nitrogen and 4\% for helium. Given the low energy needed to boil cryogens and intense local heating during a quench, monitoring stray-capacitance between insulated magnet structures or dedicated porous capacitor sensors provides a clear signal of a hot-spot forming in a magnet, as the heat will rapidly boil-off the small volume of liquid cryogen present as a film between insulator and metal structures or impregnating porous insulation materials. This technique has been implemented in various helium bath cooled HTS Bi-2212 magnet geometries. Global structural stray-capacitance monitoring demonstrated sensitivity as low as 0.1 J of deposited heat and clear deviations up to 10 seconds prior to thermal run-away. Basic plate sensors symmetrically placed on the straight sections of a racetrack coil were able to localize heater induced quenches with both response times and peak capacitance change. Capacitance monitoring is an effective diagnostic tool for liquid cryogen systems, particularly with HTS magnets low-normal zone propagation and broader transition which make conventional quench detection more challenging \cite{davis-1,15-MM,davis-3}.

 	\begin{figure}[H]
		\centering
		
		\includegraphics[width=5in]{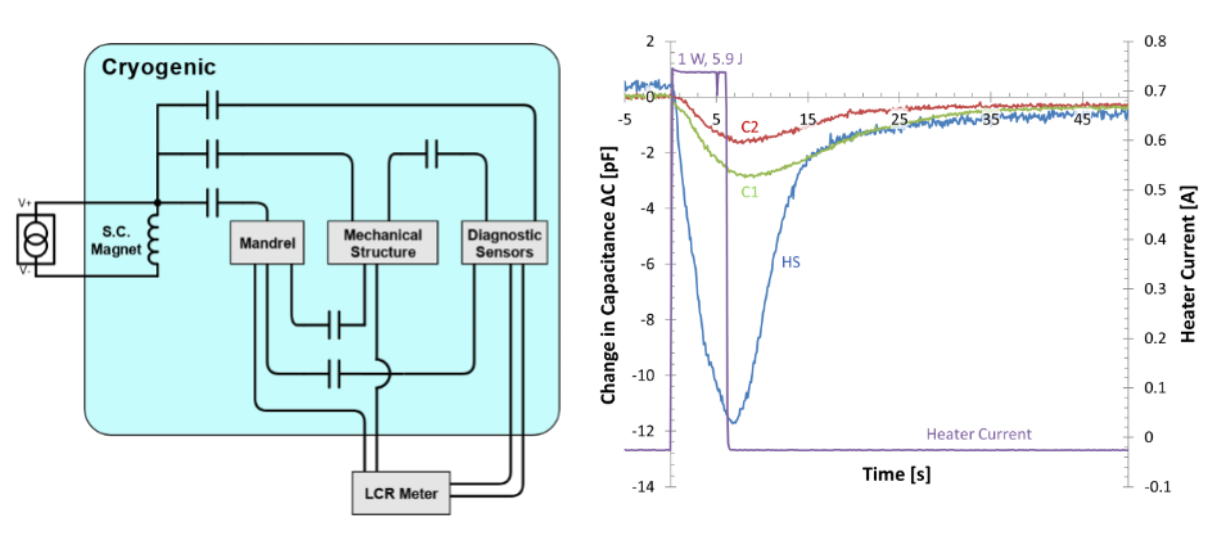}		
		\caption{(Left) A generalized circuit representation of stray capacitance monitoring for a superconducting magnet system. (Right) Typical capacitance response from RC6, a single racetrack coil encased in a steel structure, for local strip sensors C1 (green) \& C2 (red), and structural sensor HS (blue) to a heater current pulse (purple) in liquid helium \cite{davis-3}.}
		\label{Fig:davis_1}
	\end{figure}

\subsection*{Fiberoptic diagnostics}

Over the past 20 years, fiber optic sensors have been identified as a promising diagnostic tool for superconducting magnets \cite{1-Maria}. Fiber optics are inexpensive and well-established technologies employed in several industrial sectors. The working principle is simple; the spectral shift observed in the fiber can be directly connected to strain and temperature variations \cite{16-MM}. The employment of fibers presents several advantages: fibers for example are not sensitive to electromagnetic fields. Those sensors can be divided in two main categories: discrete sensors based on FBG or distributed sensors based on Rayleigh, Raman or Brillouin backscattering. FBG sensors are currently used at CERN to monitor strain variations over the entire lifetime of MQXFB magnets including cool-down, energization and quench \cite{3-Maria,4-Maria}. 

Distributed fibers based on Rayleigh back-scattering appear to be an especially promising diagnostic tool. These sensors could provide several advantages over traditional techniques for detecting normal zones \cite{5-Maria}. Indeed, Rayleigh-scattering based fibers were used to detect the temperature change with a high spatial resolution in the millimeter range \cite{1-Maria}. Fibers have been also integrated into a REBCO conductor architecture and demonstrated strain sensing capabilities as well as thermal perturbation detection and localization with higher spatial resolution than voltage taps \cite{14-MM,7-Maria,8-Maria}. More R\&D development will be necessary for employment of Rayleigh distributed sensors in superconducting magnets. Optimal coating materials to improve fiber sensitivity need to be identified. Spatial and temporal resolution need to be increased in addition to the length of the fiber itself.

 	\begin{figure}[H]
		\centering
		\includegraphics[width=5.25in]{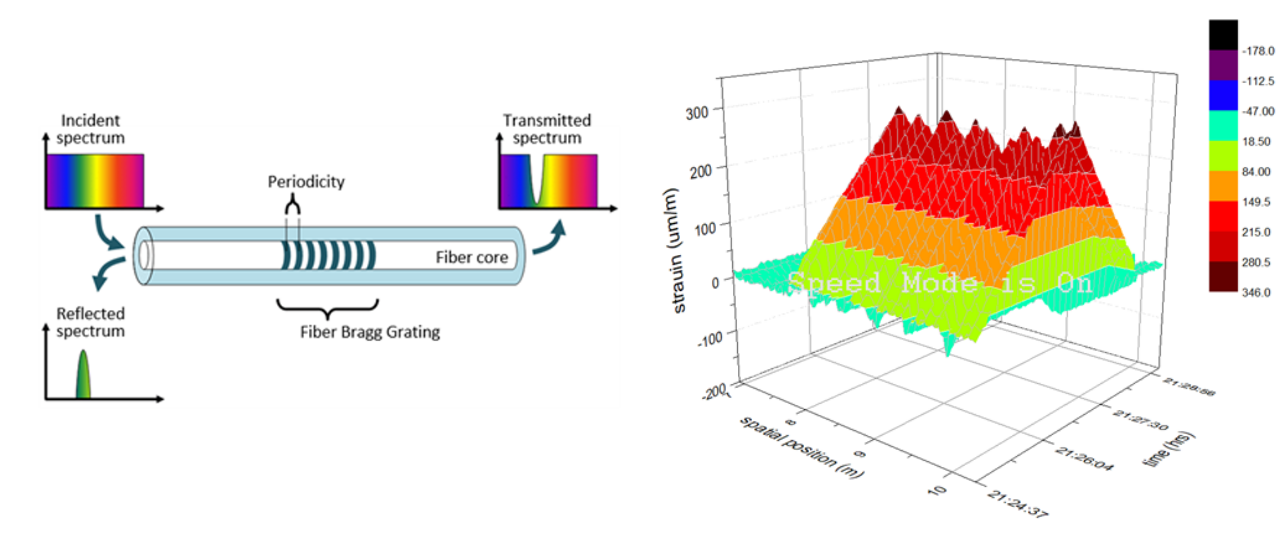}		
		\caption{(Left) Working principle of FBG sensors; modulation of the refractive index allows a single wavelength to be reflected, with strain variation proportional to the spectral shift. (Right) Example fiber measurement showing strain at six gratings during tensile loading and unloading.}
		\label{Fig:maria_1}
	\end{figure}

\subsection*{Cryogenic FPGAs and Data Acquisition}

\begin{figure}[H]
		\centering
		\includegraphics[width=4in]{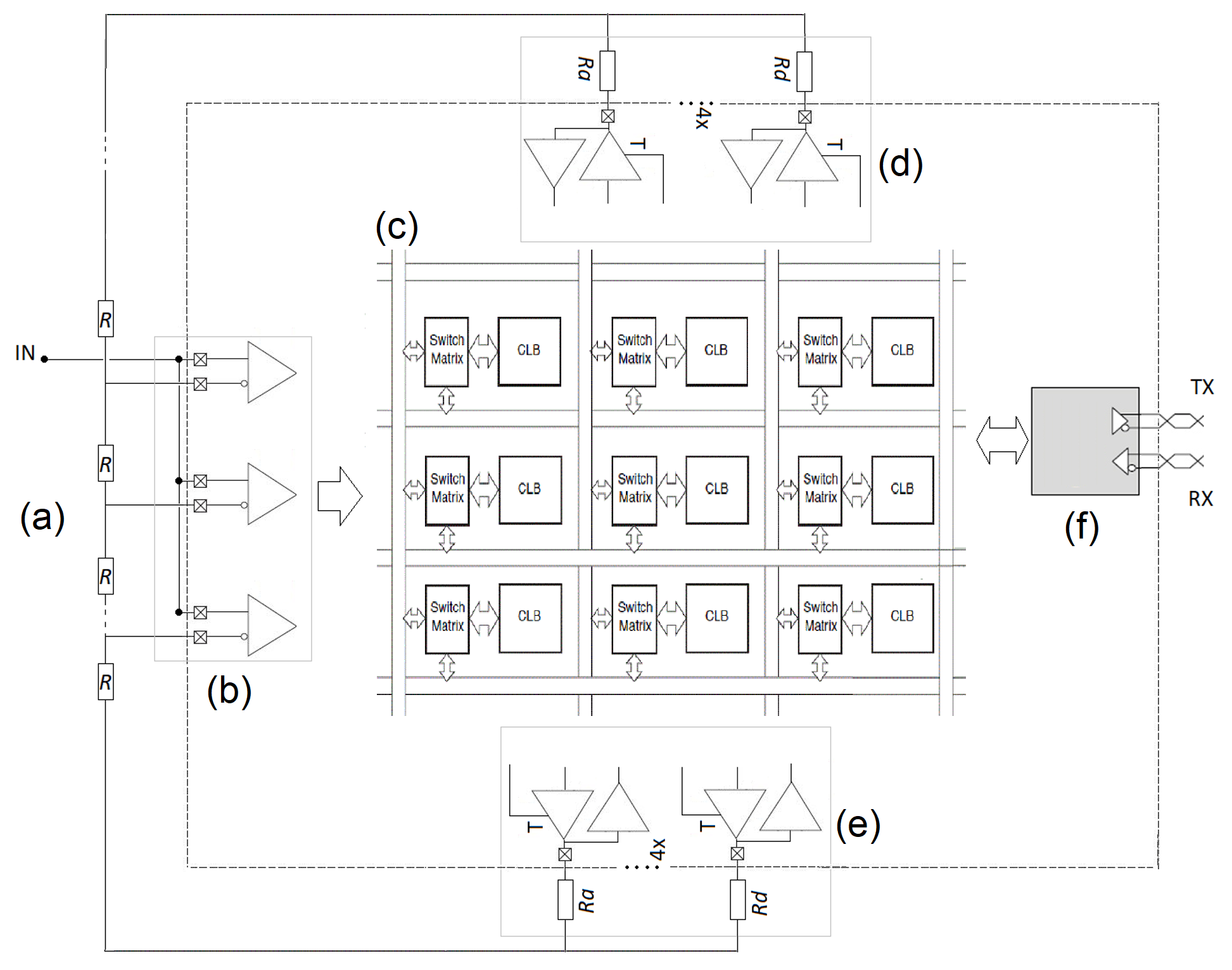}		
		\caption{FPGA based cryogenic Re-configurable Flash Folding Architecture. With this architecture it is possible to change the ADC resolution, and sampling rate on the fly.(a) Front-end resistive array, (b) comparators, (c) FPGA fabric, (d) and (e) output buffers for voltage reference, and (f) serial communication.}
		\label{Fig:marcos_1}
	\end{figure}

Data acquisition systems operating at cryogenic temperatures are growing in numbers due to the increased need of several fields of industry and research, including superconducting magnets and insertion devices. Currently, warm electronics operating relatively far from analog signal origins are used to readout and manage cryogenic sensors. Warm operation coupled with the distance that analog signals must travel degrades the signal of the probed sensor. The best way to preserve signal integrity of sensors running cold is to locate the data acquisition as close as possible to the signal source. In order to achieve this proximity, the electronic circuitry needs to operate at the same temperature as the sensing element. FPGA devices capable of liquid Helium (LHe) operation are a powerful tool to achieve this goal \cite{1-MT,2-MT}. The cold operation of such device means that we can now implement complex electronics at deep cryogenics temperatures dramatically reducing the length of analog lines coming from the cold system, whatever the signal source may be. This is especially important for systems with many analog channels as it simplifies the cryostat design, eliminates the need for analog buffers, and allows for a higher channel density. The versatility provided by FPGAs gives designers the possibility to essentially implement a wide range of digital circuits and a few analog circuits. Some especially important circuits demanded by data acquisition systems that can be implemented in FPGAs include; hybrid analog to digital converters, phased-locked loop, high speed serial communications, digital signal processing, microprocessors, and memory elements. All these capabilities are available for deep cryogenic circuit design in a single device; the FPGA (Figure.~\ref{Fig:marcos_1}).

\subsection*{Data analysis with Machine Learning}

A promising approach for understanding quench origins in both LTS and HTS magnets is the search for quench precursors via anomalies in diagnostic data. Although quench precursors have been seen in post-processing, Machine Learning (ML) holds promise to detect quench precursors in real-time as a means of magnet protection. ML algorithms can be trained to learn how the magnet behaves during typical operation, and identify anomalies in monitored data at high speed. The use of unsupervised algorithms, such as auto-encoders, has already yielded promising results; a prototype off-line algorithm was able to find anomalies in 77\% of the data using acoustic measurements \cite{1-ML}. There are opportunities for further development and alternative methodologies in this area, such as convolutional deep neural networks, recurrent neural network architectures (including long short-term memory layers \cite{2-ML}), or attention \cite{3-ML} or graph-based architectures in order to build more robust and reliable real-time quench precursor detection systems.

\section*{Conclusion}

A synergistic analysis of data acquired by these diverse diagnostic techniques will bring us closer to answering key technical questions that define SC magnet performance. It is an ample and comprehensive program with the aim of developing an integrated system of hardware and software solutions applicable not only to the U.S. MDP SC magnets, but to any other SC accelerator magnet and also magnets for test facilities. The effort should extend well beyond MDP and engage instrumentation experts across national and international labs. 

\clearpage
\bibliographystyle{IEEEtran}
\bibliography{references}
\end{document}